\documentclass[journal=jpccck,manuscript=article]{achemso}

\setkeys{acs}{articletitle=true}

\usepackage{graphicx}
\usepackage{subfigure}
\usepackage{multirow}
\usepackage{chemarr}
\usepackage{amssymb}
\usepackage{xspace}
\usepackage{color}

\author{Padmabati Mondal} \altaffiliation{Current address: Department
  of Chemistry and Center for Atomic, Molecular and Optical Sciences
  and Technologies, Indian Institute of Science Education and Research
  (IISER) Tirupati, Karakambadi Road, Mangalam, Tirupati-517507,
  Andhra Pradesh, India.}
\email{padmabati.mondal@iisertirupati.ac.in}

\author{Pierre-Andr\'{e} Cazade} \altaffiliation{Current address:
  Bernal Institute, University of Limerick, Plassey Park Road,
  co. Limerick, Castletroy, Ireland}

\author{Akshaya K. Das} \altaffiliation{University of Calfornia, Berkeley}

\author{Tristan Bereau} \altaffiliation{Van 't Hoff Institute for
  Molecular Sciences and Informatics Institute, University of
  Amsterdam, Amsterdam 1098 XH, The Netherlands and
  Max-Planck-Institut f\"ur Polymerforschung, Ackermannweg 10, 55128
  Mainz, Germany}

\author{Markus Meuwly} \affiliation[University of Basel]{Department of
  Chemistry, University of Basel, Klingelbergstrasse 80, 4056 Basel,
  Switzerland} \alsoaffiliation[Brown University]{Department of
  Chemistry, Brown University, Providence/RI, USA}
\email{m.meuwly@unibas.ch}

\title{Multipolar Force Fields for Amide-I Spectroscopy from
  Conformational Dynamics of the Alanine-Trimer}

\keywords{2DIR, NMA, Multipoles, MD, Trialanine, Conformational dynamics}

\begin{document}
\date{\today}

\begin{abstract}
The dynamics and spectroscopy of N-methyl-acetamide (NMA) and
trialanine in solution is characterized from molecular dynamics (MD)
simulations using different energy functions, including a conventional
point charge (PC)-based force field, one based on a multipolar (MTP)
representation of the electrostatics, and a semiempirical DFT
method. For the 1-d infrared spectra, the frequency splitting between
the two amide-I groups is 10 cm$^{-1}$ from the PC, 13 cm$^{-1}$ from
the MTP, and 47 cm$^{-1}$ from SCC-DFTB simulations, compared with 25
cm$^{-1}$ from experiment. The frequency trajectory required for
determining the frequency fluctuation correlation function (FFCF) is
determined from individual (INM) and full normal mode (FNM) analyses
of the amide-I vibrations. The spectroscopy, time-zero magnitude of
the FFCF $C(t=0)$, and the static component $\Delta_0^2$ from
simulations using MTP and analysis based on FNM are all consistent
with experiments for (Ala)$_3$. Contrary to that, for the analysis
excluding mode-mode coupling (INM) the FFCF decays to zero too rapidly
and for simulations with a PC-based force field the $\Delta_0^2$ is
too small by a factor of two compared with experiments. Simulations
with SCC-DFTB agree better with experiment for these observables than
those from PC-based simulations. The conformational ensemble sampled
from simulations using PCs is consistent with the literature
(including P$_{\rm II}$, $\beta$, $\alpha_{\rm R}$, and $\alpha_{\rm
  L}$), whereas that covered by the MTP-based simulations is dominated
by P$_{\rm II}$ with some contributions from $\beta$, $\alpha_{\rm
  R}$. This agrees with and confirms recently reported,
Bayesian-refined populations based on 1-dimensional infrared
experiments. Full normal mode analysis together with a MTP
representation provides a meaningful model to correctly describe the
dynamics of hydrated trialanine.
\end{abstract}

\section{Introduction}
Ultrafast Infrared (IR) spectroscopy is a powerful tool to
characterize the solvent dynamics around chromophores on the pico- and
sub-picosecond time scale. It has also been proven to be a promising
tool for studying the structure and dynamics of proteins, including
protein-folding and protein-ligand
binding.\cite{Tokmakoff.amide.2008,Getahun.cn.jacs.2003,Bagchi.cnprot2dir.jpcb.2012,Xu.vibstarkhalr2.bc.2011,Mondal.pccp.2017,hamm.aha,sharon.vibstarkcn.jacs.2013}
The amide-I mode is suitable to probe the structural dynamics and the
conformational ensemble of a solvated molecule, peptide, or
protein.\cite{Tokmakoff.amide.2008,MM.insulin:2020} Other suitable
vibrational labels\cite{gai:2011,hamm.rev:2015} that absorb in the
spectroscopic window between $\sim 1700$ and $\sim 2800$ cm$^{-1}$ are
cyanophenylalanine\cite{thielges:2015}, nitrile-derivatized amino
acids,\cite{gai:2003} the sulfhydryl band of
cysteines,\cite{hamm:2008} deuterated carbons,\cite{romesberg:2011}
non-natural labels consisting of metal-tricarbonyl modified with a
-(CH$_2$)$_{\rm n}$- linker,\cite{zanni:2013} nitrile
labels,\cite{fayer.ribo:2012} cyano\cite{romesberg.cn:2011} and
SCN\cite{bredenbeck:2014} groups, or cyanamide.\cite{cho:2018}
Contrary to these other probes the amide-I band characterizes the
inherent dynamics of the system because it does not require mutation
or chemical modification of the molecule considered.  \\

\noindent
N-methyl acetamide (NMA) is a typical model system for
experimental\cite{Hamm.2dirbook,Hamm.jpcb.1998,Zanni.jcp.2001,
  hamm:2001,skinner.nma:2011}, and
computational\cite{Cazade2014,Chandra2015,Gaigeot,Cazade2012} studies
because it is also the fundamental building block to study longer
peptides and proteins. In going from a mono- to a poly-peptide one
essentially moves from NMA to alanine dipeptide, to trialanine and to
larger alanine chains. Therefore, to develop and validate force fields
for the amide probe and to apply them to larger polypeptide chains,
starting from NMA is a meaningful choice. This also allows one to
assess the transferability of the force fields from NMA by using them
for polypeptides and comparing the results with experimental data.\\

\noindent
Two-dimensional infrared (2D-IR) spectroscopy provides quantitative
information about the solvent structure and dynamics surrounding a
solute.\cite{hamm:2015} Such techniques are particularly useful to
measure the fast (picosecond) dynamics in condensed-phase systems. The
coupling between inter- and intramolecular degrees of freedom - such
as the hydrogen bonding network in solution, or the conformational
dynamics of biological macromolecules - can be investigated by
monitoring the fluctuation of a fundamental vibrational frequency,
which is the amide-I mode in the present work. Computationally, this
information is accessible from either instantaneous normal modes
(NM),\cite{StrattJCP1994,Cazade2014,Mondal.pccp.2017} the solution of
a reduced-dimensional nuclear Schr\"odinger
equation,\cite{MM.n3:2019,Koner.jcp.2020} or from spectroscopic
maps.\cite{skinner-map-JPCB2011} This frequency trajectory
($\omega(t)$ or $\nu(t)$ for harmonic or anharmonic vibrations,
respectively) is then used to determine the frequency fluctuation
correlation function which can be directly compared with experimental
measurements.\\

\noindent
The linear and non-linear vibrational spectroscopy and conformational
dynamics of trialanine in solution has been investigated from both,
experiments and
computations.\cite{woutersen:2000,Woutersen,Hamm.pnas.2001,Schweitzer-Stenner.jacs.2001,Woutersen:2001,Mu.jpcb.2002,Graf2007,Gorbunov2007,Oh2010,Xiao2014,Tokmakoff2018}
Computationally, a quantum-classical description of the amide-I
vibrational spectrum of trialanine in D$_2$O probed different
approximations typically made in determining the vibrational
lineshapes. \cite{Gorbunov2007} A combined experimental and molecular
dynamical study using non-linear time-resolved spectroscopy on
trialanine found conformational heterogeneity of the
peptide.\cite{Woutersen} Peptide conformational ensembles were also
studied for trialanine using two-dimensional IR and NMR
spectroscopies.\cite{Tokmakoff2018,Xiao2014,Oh2010} Two-dimensional IR
studies probed the subpicosecond dynamics\cite{Hamm.pnas.2001} and
with isotopically labelled (Ala)$_3$ the dipole-dipole coupling
strength was determined.\cite{Woutersen:2001} Including such couplings
is often done in models based on spectroscopic maps. In the present
work, NMs are determined from ``independent normal modes'' (INM) and
from a ``full normal mode'' (FNM) analysis which allows coupling of
two or several amide-I modes.\\

\noindent
The present work is structured as follows. First, the methods used are
introduced. This is followed by an analysis the spectroscopy and
dynamics of solvated, deuterated N-methyl-acetamide with a flexible
solute. Next, the spectroscopy and structural dynamics of trialanine
are discussed. Finally, conclusions are drawn.\\

\section{Computational methods}

\subsection{Molecular Dynamics Simulations}
\begin{figure}[!h]
\includegraphics[width=10cm]{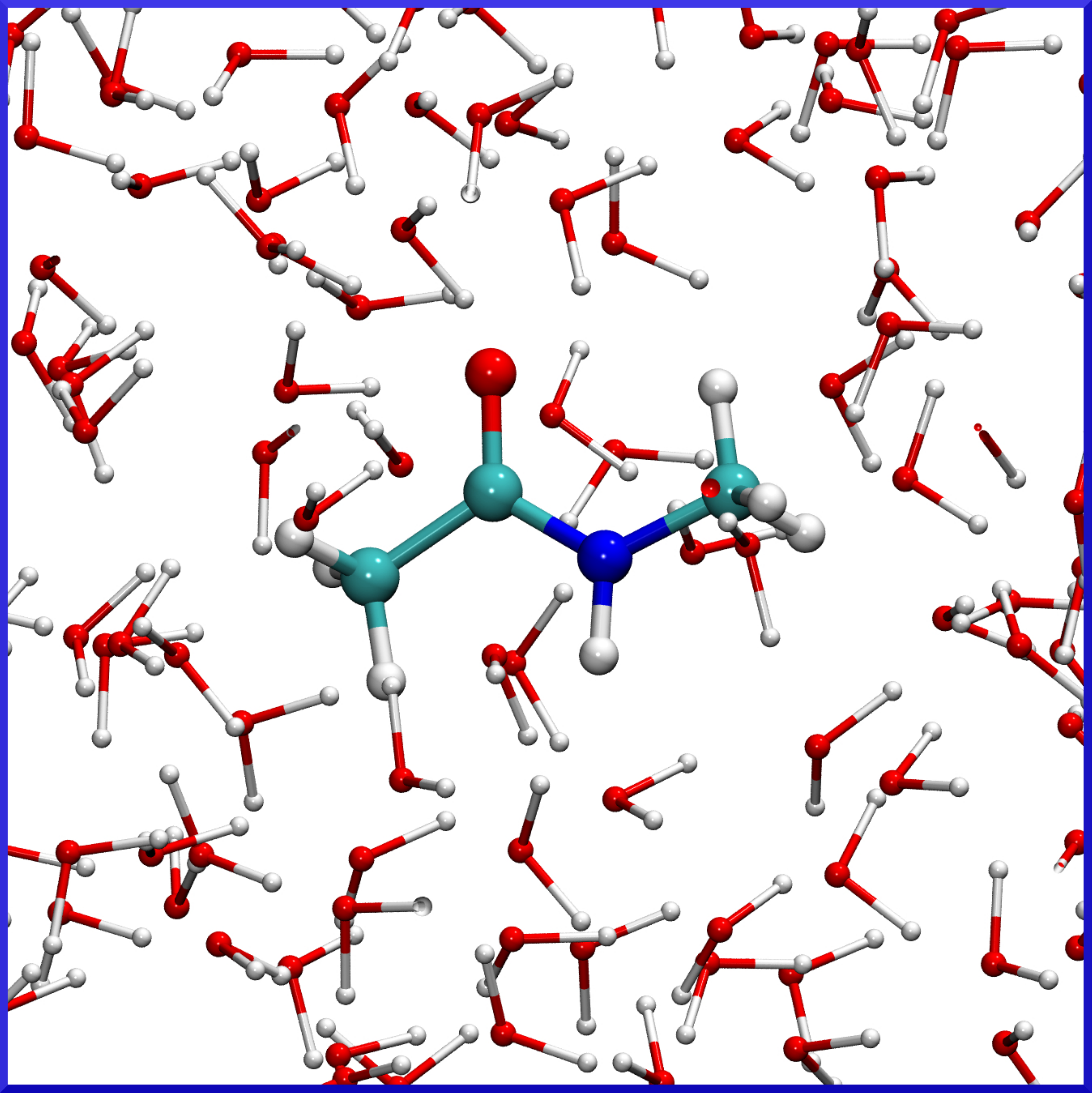}
\caption{N-methylacetamide (NMAD) solvated in a periodic cubic box of
  D$_2$O of size $30^3$ \AA\/$^3$. Atom color code: Carbon (cyan),
  Oxygen (red), Nitrogen (blue) and Hydrogen (white).}
\label{fig:system}
\end{figure}

Molecular Dynamics (MD) simulations were carried out for N-deuterated
N-methylacetamide (NMAD, see Figure \ref{fig:system}) and trialanine
(Ala)$_3$ in a periodic cubic box of deuterated
TIP3P\cite{Jorgensen83p926} water molecules. The box size was $30^3$
\AA\/$^3$ and the system consisted of one solute molecule surrounded
by 882 water molecules (for NMAD) and 795 water molecules (for
(Ala)$_3$), respectively. (Ala)$_3$ was fully deuterated and the
positively charged species (i.e. ``cationic'' with ND$_3^+$ and COOD
termini) was investigated.\cite{woutersen:2000,Tokmakoff2018} To
neutralize the simulation system, one chloride ion was added and
constrained in one corner of the simulation system during MD
simulations.\\

\noindent
All MD simulations were performed with the CHARMM program\cite{charmm}
with provision for multipolar
interactions.\cite{bereau2013leveraging,Kramer12p1673} Parameters for
NMA are based on CGenFF\cite{vanommeslaeghe2010charmm} unless stated
otherwise and described in more detail in
Ref.~\cite{Cazade2014}. Electrostatic interactions were treated using
Particle-Mesh Ewald (PME)\cite{pmeDarden_jcp_1993} with a grid-size
spacing of 1 \AA\/, characteristic reciprocal length $\kappa=
0.43$~\AA$^{-1}$, and interpolation order 4 for long-range
electrostatics. For the Lennard-Jones (LJ) interactions a 12 \AA\/
cut-off and 10 \AA\/ switching were used. The simulations were
performed at $T = 300$ K and all bonds involving hydrogen atoms were
constrained via the SHAKE algorithm.~\cite{shake-gunsteren} The
timestep was $\Delta t = 0.5$~fs and snapshots were recorded every 10
time steps.\\

\noindent
Mixed QM/MM were carried out using Self-consistent charge density
functional tight-binding (SCC-DFTB)\cite{mio98} as implemented in
CHARMM.\cite{qmmmsccdftb} In these simulations, the entire solute
(NMAD or (Ala)$_3$) was treated with SCC-DFTB whereas all water
molecules and the ion (for the solvated (Ala)$_3$ system) were treated
by MM. First the system was minimized and heated to 300 K. A {\it NVT}
simulation was carried out at 300 K using the velocity Verlet
integrator with a (shorter) time step of $\Delta t = 0.25$ fs for 5
ns. Again, all bonds involving hydrogen atoms were constrained using
SHAKE\cite{shake-gunsteren} and the treatment of the nonbonded
interactions was that same as that for the PC and MTP simulations
described above. \\

\subsection{Force fields for flexible NMA}
Two different electrostatic models for NMA are used in this work. The
first one uses point charges (PCs) based on the CGenFF force
field. The second model is the multipolar MTPW representation
including atomic multipoles up to quadrupoles on heavy atoms for the
entire NMAD molecule.\cite{Cazade2014}. The force field parameters for
the CO bond are based on {\it ab initio} calculations at the
MP2/6-31G$^{**}$ level and are readjusted to reproduce the gas phase
amide-I frequency. The Morse parameters are $D_e = 141.67$ kcal
mol$^{-1}$, $\beta = 2.11$ \AA$^{-1}$ and $r_{\rm eq} = 1.23$ \AA\/.\\

\noindent
The parametrization for (Ala)$_3$ uses the CGenFF force
field\cite{cgenff} except for the CO-stretch potential which is the same
Morse function used for the -CO group of NMAD and the multipoles on the C-terminal CO atoms as well as outer and
central [CONH] atoms which were also used for [CONH] group of NMAD.\\

\subsection{The Frequency Fluctuation Correlation Function and 1D infrared spectrum}
The FFCF, $C(t)$, is obtained from the frequency trajectory
$\omega(t)$ according to:
\begin{equation}
  \label{freqcorr}
\begin{split}
C\left( t \right)&=\left\langle
\delta\omega\left(t_0\right)\delta\omega\left(t_0 + t
\right)\right\rangle_{t_0}\\ &=\left\langle
\left(\omega\left(t_0\right)-\overline{\omega}\right)\left(\omega\left(t_0
+ t \right)-\overline{\omega}\right)\right\rangle_{t_0}.\\
\end{split}
\end{equation}
Here, $\omega(t_0)$ is the instantaneous frequency at time $t_0$ and
$\overline{\omega}$ is the average frequency and thereby $\delta\omega
(t_0)$ refers to the frequency fluctuation at time $t_0$. The
instantaneous frequencies $\omega(t)$ are obtained from normal mode
(NM) calculations. For each snapshot of the trajectory, the structure
of the solute (here NMAD and (Ala)$_3$) is minimized while keeping the
solvent frozen. Frequencies are calculated using two different
approaches referred to as ``full NM'' (FNM) and ``independent NM''
(INM) analysis methods. For FNM the normal mode analysis is carried
out for the entire solute. Such an approach includes both, the
frequencies of the labels (``site energies'') and the couplings
between them. On the other hand, INM refers to the normal mode
analysis of the independent amide modes of trialanine while keeping
everything except the [CONH] group fixed and therefore neglects the
couplings between the spectroscopic labels. This approach is
computationally more efficient than scanning along the normal mode and
solving the 1- or even 3-dimensional nuclear Schr\"odinger
equation.\cite{MM.n3:2019,MM.insulin:2020,MM.azidolys:2021}\\

\noindent
The analysis adopted here is also reminiscent of instantaneous normal
modes (NM) which have been shown to perform well for the short-time
dynamics in condensed
phase.\cite{stratt:inm:1990,stratt:inm:1992,Bastida2010,stratt:inm:2013}
Furthermore, a direct comparison between instantaneous normal modes,
scanning (``scan'') along the local and normal mode and map-based
frequency trajectories has been recently presented and found that
``NM'' and ``scan'' yield comparable FFCFs and 1d-lineshapes derived
from them.\cite{MM.insulin:2020}\\

\noindent
The 1D and 2D response functions can be determined from the lineshape
function $g(t)$,\cite{Hamm.2dirbook,schmidt2007} which is related to
the FFCF through
\begin{equation}
g(t)=\int_0^t \int_0^{\tau'} d\tau'
d\tau'' \left \langle \delta\omega(\tau'') \delta\omega(0) \right\rangle.
\label{eq:gt}
\end{equation}
Depending on whether or not the FFCF is fit to a parametrized form,
the double integration can be carried out in closed form or needs to
be done numerically. In the present case, the functional form fitted
to is
\begin{equation}
C(t)= \sum_{i=1}^{n} a_i \exp{(-t / \tau_i)} +\Delta_0^2
\label{eq:fit.ffcf}
\end{equation}
with amplitudes $a_i$ and decay times $\tau_i$ as fitting parameters
and $n_{\rm max} =2$ or 3 to make direct comparison with earlier work
on (Ala)$_3$.\cite{stock.nma.jcp.2002} The $a_i$ and $\tau_i$ are
amplitudes and relaxation times, respectively, and $\Delta_0^2$ is the
static component which can differ from 0 for situations in which
processes occurring on longer time scales have not equilibrated on the
time scales of the relaxation times $\tau_i$.\\

\section{Results}

\subsection{Spectroscopy of N-methylacetamide}
To validate the energy functions and analysis techniques used
subsequently for (Ala)$_3$ first the spectroscopy of NMAD in D$_2$O
from MD simulations with PCs and MTPs for flexible solute were
considered. In addition, QM(SCC-DFTB)/MM simulations were also carried
out. For each of the three cases, $10^6$ snapshots from a 5 ns long
trajectory were analyzed. For every snapshot the frequency,
$\omega(t)$, was obtained from an instantaneous normal mode
analysis. From this, the FFCFs were determined and fitted to
multi-exponential decay functions along with a static component
($\Delta_0^2$) according to Eq. \ref{eq:fit.ffcf} with $n_{\rm max} =
2$ or $3$. \\

\begin{figure}[!h]
\includegraphics[scale=0.5]{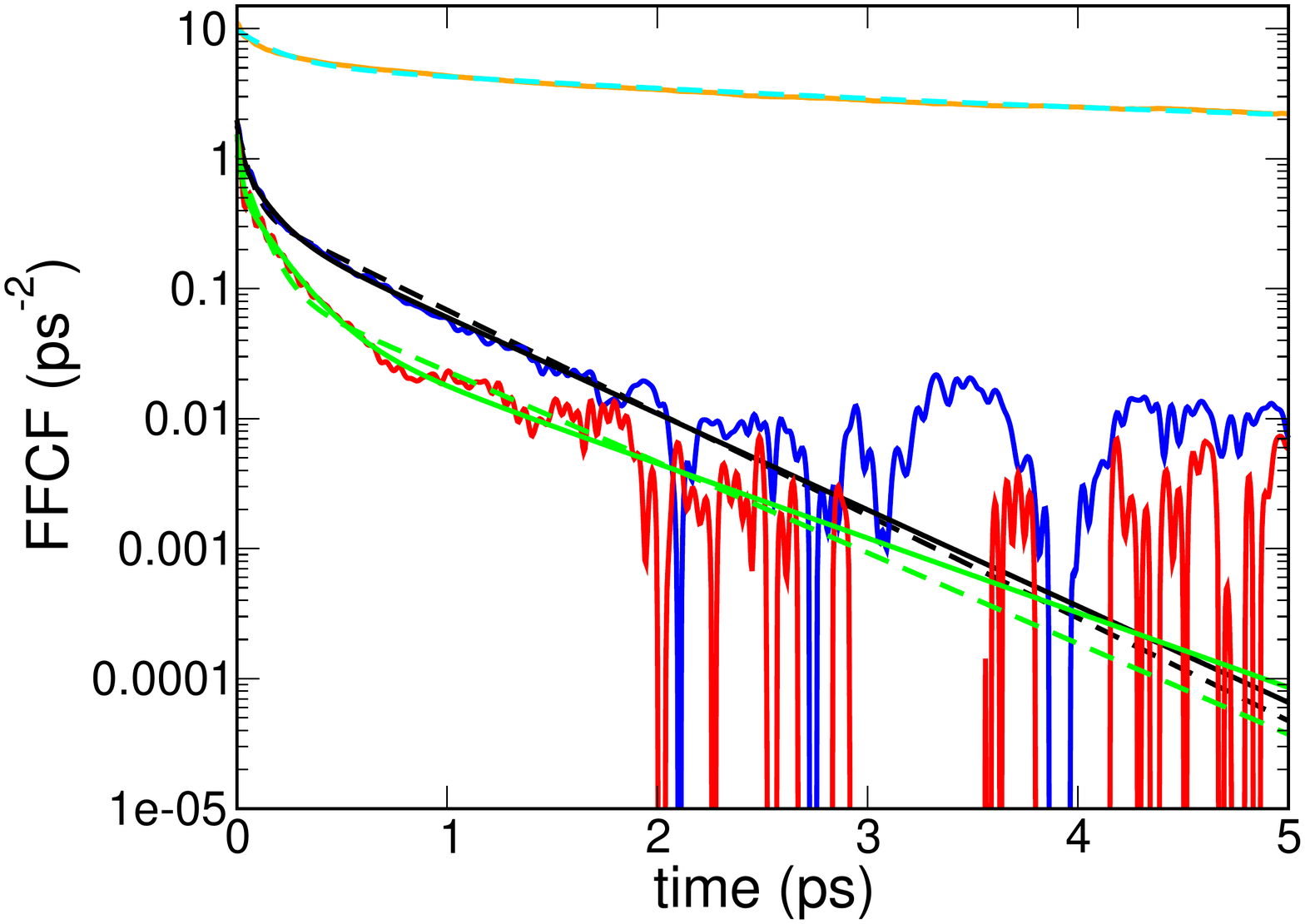}
\caption{FFCFs for NMAD in D$_2$O (TIP3P) for simulations with PC
  (red), MTP (blue), and SCC-DFTB (orange) with flexible NMA. Green,
  black and cyan lines are fits to Eq.  \ref{eq:fit.ffcf} for the
  FFCFs from simulations with PC, MTP, and SCC-DFTB,
  respectively. Dashed lines are for fits using $n_{\rm max} = 2$ and
  solid lines for fits with $n_{\rm max} = 3$ in
  Eq. \ref{eq:fit.ffcf}.}
\label{fig:fig1}
\end{figure}

\noindent
Figure \ref{fig:fig1} shows the FFCF for NMAD in D$_2$O for the
simulations with PC (red), MTP (blue), and SCC-DFTB (orange)
models. The fits, using two or three time scales, respectively, are
the dashed and solid green, black and cyan lines. The fitting
parameters for the FFCFs are summarized in
Table~\ref{tab:tab1}. Figure \ref{fig:fig1} shows that for the PC, MTP
and SCC-DFTB, two time scales are sufficient to represent the FFCF. Also, for the two force
field models the FFCFs decay to zero on the $\sim 10$ ps time scale
whereas that from the SCC-DFTB simulations has a static component of
$\Delta_0^2 = 1.3$ ps$^{-2}$.\\

\noindent
The short time decay $\tau_1$ for the PC and MTP models range from
0.02 ps to 0.05 ps, consistent with experiments (between 0.01 and 0.1
ps).\cite{Woutersen,Decamp} Contrary to that, simulations with
SCC-DFTB yield $\tau_1 = 0.18$ ps which is at least a factor of two
slower compared with what has been reported from experiments. The long
time scale, $\tau_3$, ranges from 0.55 ps to 0.62 ps, compared with
1.0 ps and 1.6 ps from the experiments.\cite{Woutersen,Decamp} Earlier
MD simulations reported $\tau_3 = 0.66$ ps.\cite{Decamp} The SCC-DFTB
simulations find a long time scale $\tau_3 = 3.2$ ps which is longer
than any of the experiments. It is also worthwhile to note that a
two-time scale fit of the FFCF to the frequencies from the MTP
simulation is sufficient and assuming three time scales does not
provide additional information. This is also found from the
experiments.\cite{Decamp} The fits with only two time scales are
preferred as with every additional time scale a new process is
associated. For water the sub-picosecond time scale has been
associated with partial water reorientation whereas the process on the
ps time scale is considered to involve full water
reorientation.\cite{laage:2011} \\

\begin{table}[H]
\centering
\begin{tabular*}{\linewidth}{@{\extracolsep{\fill}} lccccccc}
\hline
Model & $a_1$ [ps$^{-2}$] & $\tau_1$ [ps] & $a_2$ [ps$^{-2}$] & $\tau_2$ [ps] & $a_3$ [ps$^{-2}$] & $\tau_3$ [ps] &$\Delta_0^2$[ps$^{-2}$]  \\
\hline
PC & 0.470 & 0.019 &  0.597 & 0.075  & 0.085 & 0.588 & --\\
PC (bi-exp) & 0.951 & 0.080 & -- & -- & 0.115 & 0.622 & --\\
MTP & 0.942 & 0.019 & 0.707 & 0.110 & 0.330 & 0.587& -- \\
MTP (bi-exp) & 1.361 & 0.049 & - & - & 0.418 & 0.550 &-- \\
\hline
SCC-DFTB & 4.488 &  0.183 & -- & -- & 3.982 & 3.202 & 1.344 \\
\hline
sim. \cite{Decamp} &  &  0.06 & && & 0.66 \\
exp. \cite{Woutersen} &  &  (0.05-0.1) && & & 1.6 \\
exp. \cite{Decamp} &  &  0.01 & & && 1.0 \\
\hline
\end{tabular*}
\caption{
    Parameters of tri/bi-exponential fit (Eqn. \ref{eq:fit.ffcf})
  for MTP and PC models and bi-exponential plus static component fit
  (Eqn. \ref{eq:fit.ffcf}) for SCC-DFTB calculations for the FFCFs of
  the carbonyl group of NMAD in D$_2$O for different models.}
\label{tab:tab1}
\end{table}

\begin{figure}[H]
\includegraphics[width=15cm]{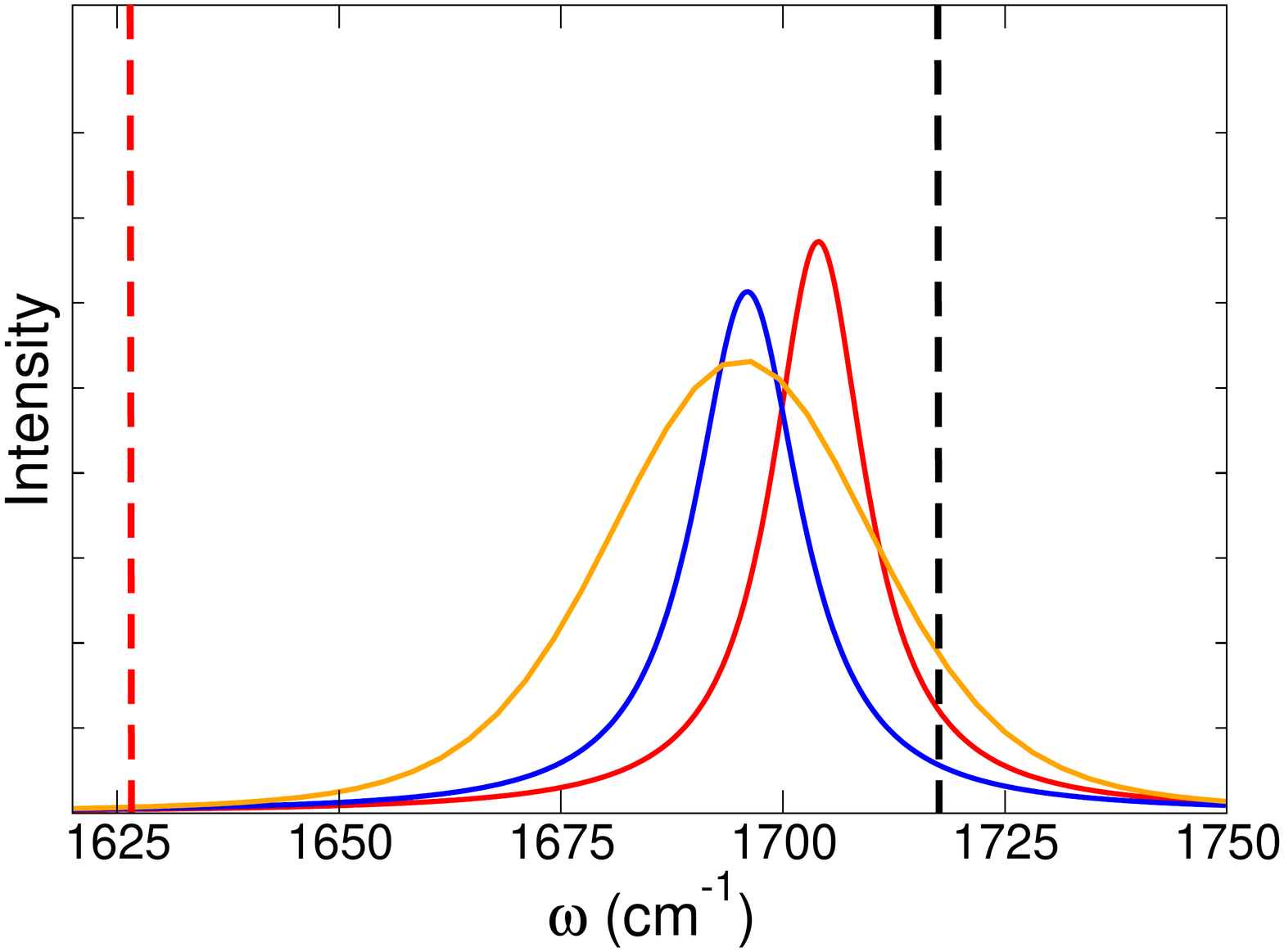}
\caption{1D absorption spectra of NMAD in D$_2$O in the region of the
  amide-I mode from simulations with the PC (red), MTP (blue), and the
  SCC-DFTB (orange) models. Experimental peaks for amide I mode of
  NMAD in D$_2$O and in the gas phase are shown as red and black
  dashed vertical line, respectively. The experimental solvent-induced
  red shift is 85 cm$^{-1}$.\cite{keiderling.nmair:2001,Jones}}
\label{fig:nmals}
\end{figure}

\noindent
The 1D absorption spectra are calculated from the analytical
integration\cite{facn,Mondal.pccp.2017} of the lineshape function (see
Eq. \ref{eq:gt}) with the FFCF ($C(t)$) fit to
Eq. \ref{eq:fit.ffcf}. A phenomenological broadening for the amide-I
vibration consistent with a lifetime of 0.45\,ps was
used.\cite{Zanni.jcp.2001} The maxima of the 1D lineshape for NMAD in
D$_2$O for PC, MTP and SCC-DFTB models are 1705 cm$^{-1}$, 1695
cm$^{-1}$ and 1695 cm$^{-1}$, respectively, see Figure
\ref{fig:nmals}. The gas phase frequency for amide mode of NMAD is
1717 cm$^{-1}$ and solvent induced red-shift for PC, MTP and SCC-DFTB
models are 12 cm$^{-1}$, 22 cm$^{-1}$ and 22 cm$^{-1}$,
respectively. This compares with an experimental solvent induced
red-shift of 85 cm$^{-1}$.\cite{keiderling.nmair:2001} The full width
at half maximum (FWHM) of the calculated 1D absorption spectra (Figure
\ref{fig:nmals}) for NMAD in D$_2$O using the PC, MTP, and SCC-DFTB
models are 12.5 cm$^{-1}$, 14 cm$^{-1}$, and 35 cm$^{-1}$, compared
with $\sim 20$ cm$^{-1}$ from experiments.\cite{stock.nma.jcp.2002}\\

\noindent
In summary, the PC and MTP models correctly capture the short and long
time scales compared with experiment with the MTP model performing
somewhat better. Simulations with both force fields correctly find
that the FFCFs decay to zero on the few-picosecond time scale whereas
SCC-DFTB leads to a static component which was not found in the
experiments. For the 1d-infrared spectroscopy, all models find a
solvent-induced red shift which, however, underestimates the
experimentally reported magnitude and the FWHM from MTP is closest to
that observed experimentally.\cite{stock.nma.jcp.2002}\\

\subsection{Spectroscopy and Dynamics of Trialanine}
\begin{figure}[H]
\includegraphics[width=10cm]{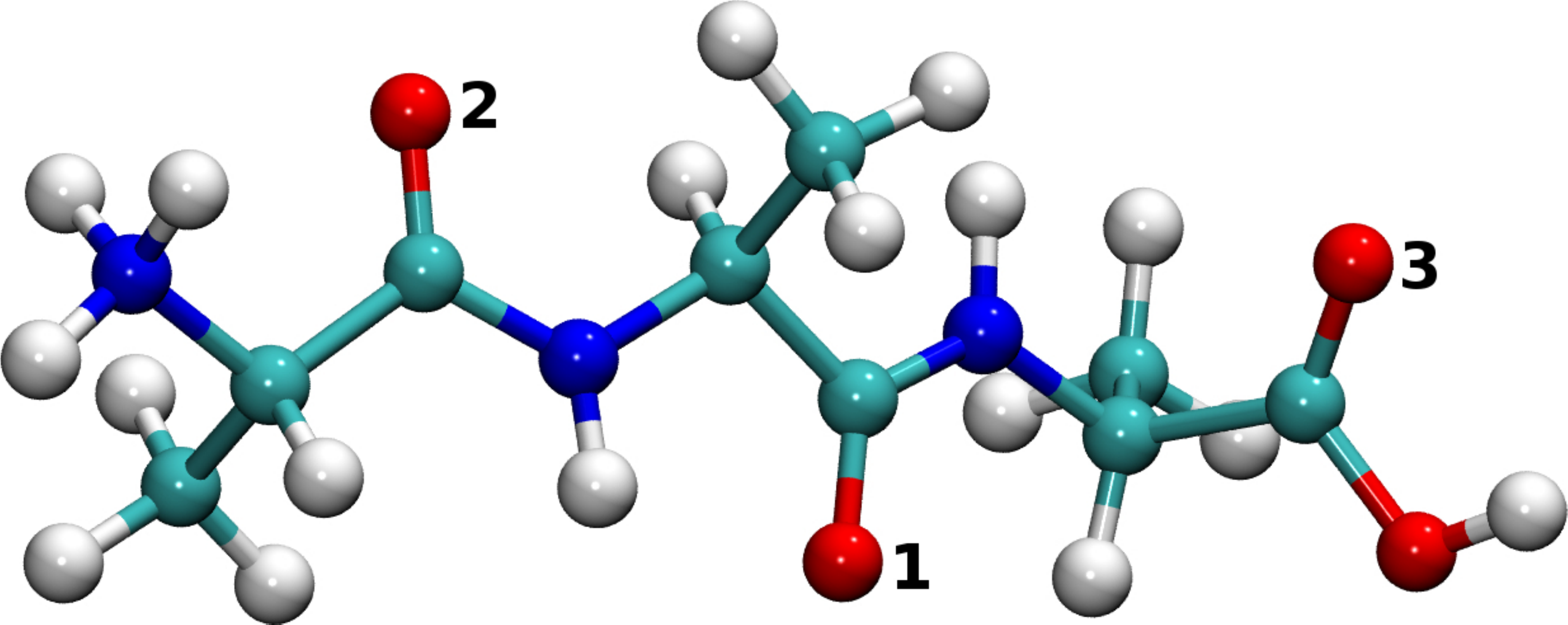} 
\caption{The structure of protonated (cationic)
  (Ala)$_3$.\cite{woutersen:2000} The central (1), outer (2) and
  carboxylic (3) -CO groups are specifically labelled. Hydrogen
  (white), oxygen (red), nitrogen (blue), and carbon (cyan) are shown
  as spheres.}
\label{fig:triala}
\end{figure}

Next, the spectroscopy and dynamics of (Ala)$_3$ (see Figure
\ref{fig:triala}) are considered. Trialanine involves two amide-I
groups (central and outer -CO) and one terminal carboxylic (COOH)
group. For each interaction model, 10 ns MD simulations were performed
for deuterated trialanine in deuterated water using PC, MTP, and
SCC-DFTB (validated for NMAD in D$_2$O). This was preceded by 1 ns of
$NPT$ equilibration and further 100 ps $NVT$ equilibration. The
-CO(OH) group of trialanine is characteristically different from the
amide -CO group. To account for this a slightly modified Morse
($\beta$) parameter (than what has been used for C=O of NMAD in
D$_2$O) is used for the C-terminal -CO group in the simulations and NM calculations.\\

\noindent
{\it Frequency Distributions:} Frequencies for the central and outer
amide as well as terminal CO(OH) are calculated using both, FNM and
INM analyses. The experimentally determined peak positions are at 1650
cm$^{-1}$, 1675 cm$^{-1}$ and 1725 cm$^{-1}$ for the central, outer
and carboxylic -CO, respectively.\cite{woutersen:2000} In the
following, results from the FNM analysis are discussed first and then
compared with those obtained from INMs, see Figure
\ref{fig:compare_full_part} and Table \ref{tab:tab2}.\\

\begin{figure}[H]
\includegraphics[width=15cm]{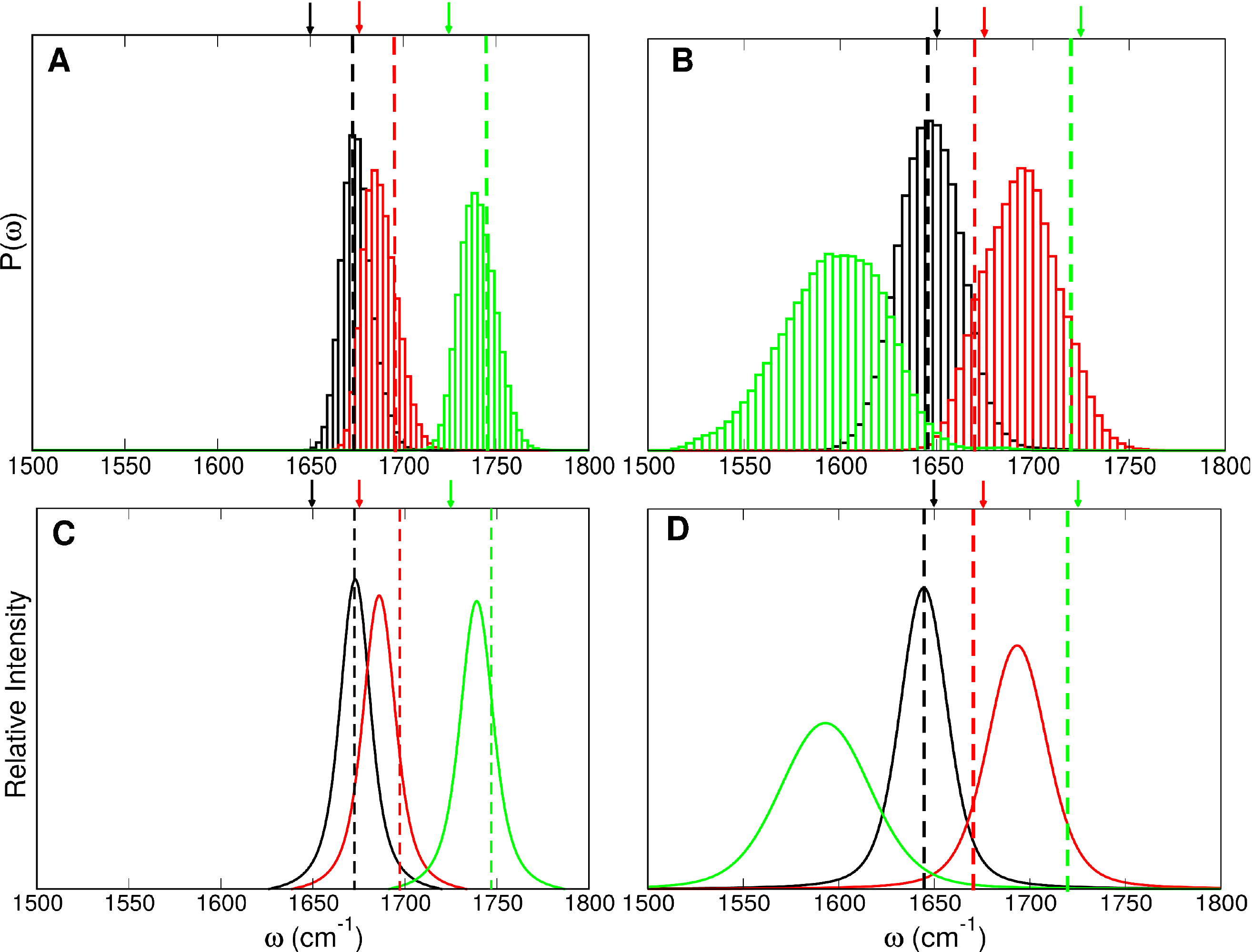}
\caption{Frequency distributions (panels A and B) and 1D absorption
  spectra (panels C and D) of each -CO moiety of trialanine for the
  central (black), outer (red), and terminal (green) -CO. Panels A and
  C correspond to results using MTP and panels B and D to those from
  SCC-DFTB simulations. The down-headed arrows indicate the
  experimental\cite{woutersen:2000} peak position of each -CO group
  whereas the vertical dashed lines with corresponding color show the
  shifted experimental peak position to best overlap with the
  simulated data of central -CO. A constant shift of 22 cm$^{-1}$
  (blue) and 5 cm$^{-1}$ (red) from the experimental spectra was
  considered to best overlap the central -CO peak for the simulations
  with MTPs and SCC-DFTB, respectively.}
\label{fig:fig3}
\end{figure}

\noindent
Figures~\ref{fig:fig3}A and B show the frequency distribution for the
central (black), outer (red) and terminal (green) carbonyl group from
simulations with the MTP and SCC-DFTB models, respectively. The
down-headed arrows of corresponding color indicate the
experimental\cite{woutersen:2000} peak positions of each -CO group and
the vertical dashed lines refer to the shifted experimental peak
position to best overlap with the simulated data for the central
-CO. This is meaningful because for the present work primarily {\it
  relative positions} of the absorption bands are of
interest. Fine-tuning of the Morse parameters to match experimental
line positions would still be possible for the PC and MTP models as an
additional refinement but is not deemed necessary here.\\

\noindent
For MTP a constant shift of 22 cm$^{-1}$ to the blue from the
experimental spectra yielded the best overlap for the central -CO
peak. The computations find a frequency of 1673\,cm$^{-1}$ for the
central -CO (black), followed by the outer -CO at 1686\,cm$^{-1}$
(red), and finally the -CO(OH) group at 1739\,cm$^{-1}$
(green). Although the same force field (MTP and Morse) was used for
the central and outer amide, the different environments experienced by
them leads to a splitting of 13 cm$^{-1}$. This sensitivity to the
environmental structure and dynamics is consistent with recent
findings for insulin monomer and dimer.\cite{MM.insulin:2020}
Nevertheless, the experimentally observed splitting of 25 cm$^{-1}$ is
still underestimated.\cite{woutersen:2000} The simulations with the PC
model also yield the correct ordering for the frequencies of the
central and outer -CO (at 1677\,cm$^{-1}$, 1687\,cm$^{-1}$ ) but the
splitting is somewhat smaller (10 cm$^{-1}$) than that from the
simulations using MTP.\\

\noindent
It is conceivable that further improvements of the
electrostatics\cite{MM.dcm:2014,MM.dcm:2017} leads to yet closer
agreement between simulations and experiments. For one,
conformationally dependent multipoles provide an even better
description of the electrostatics as has been found for isolated CO in
Mb.\cite{Nutt03p3612,Nutt04p5998,Plattner09p687,Plattner08p2505}
Furthermore, including polarizability may lead to additional
improvements.\\

\noindent
With SCC-DFTB the central, outer and terminal carbonyl peaks are at
1648 cm$^{-1}$, 1695 cm$^{-1}$ and 1598 cm $^{-1}$ (Figure
\ref{fig:fig3}B, D). A constant shift of 5 cm$^{-1}$ (red) from the
experimental spectra was considered to best overlap the central -CO
peak for the simulations with the SCC-DFTB results. Consistent with
experiment, the frequency of the outer -CO is shifted to the blue ($+
47$ cm$^{-1}$) from the central -CO by close to twice the value
reported from experiment ($+25$ cm$^{-1}$).\cite{woutersen:2000} For
the carbonyl (COOH) -CO, SCC-DFTB underestimates the frequency by 125
cm$^{-1}$ compared with experiment. This finding was reproduced from
two independent simulations. Upon visual inspection of the
trajectories it was observed that the COOH unit is typically in an
anti conformation whereas the minimum energy structure is the syn
conformer. To further validate the performance of SCC-DFTB,
simulations for (Ala)$_3$ in the gas phase using the
mio\cite{mio98,mio05} and 3ob-freq\cite{3obfreq:2013} parameter sets were
carried out. With the mio parameters, used for this study, the
frequency distributions of the central and outer -CO label are split
by $\sim 25$ cm$^{-1}$ - to be compared with a splitting of 25
cm$^{-1}$ from experiment in solution - whereas with the 3ob-freq
parametrization - which was refined for thermochemistry, geometries,
and vibrational frequencies in the gas phase - the splitting is 110
cm$^{-1}$. Hence, it is not expected that a different
parametrization will appreciably improve the findings for simulations
in solution.\\

\begin{figure}[H]
  \includegraphics[width=15cm]{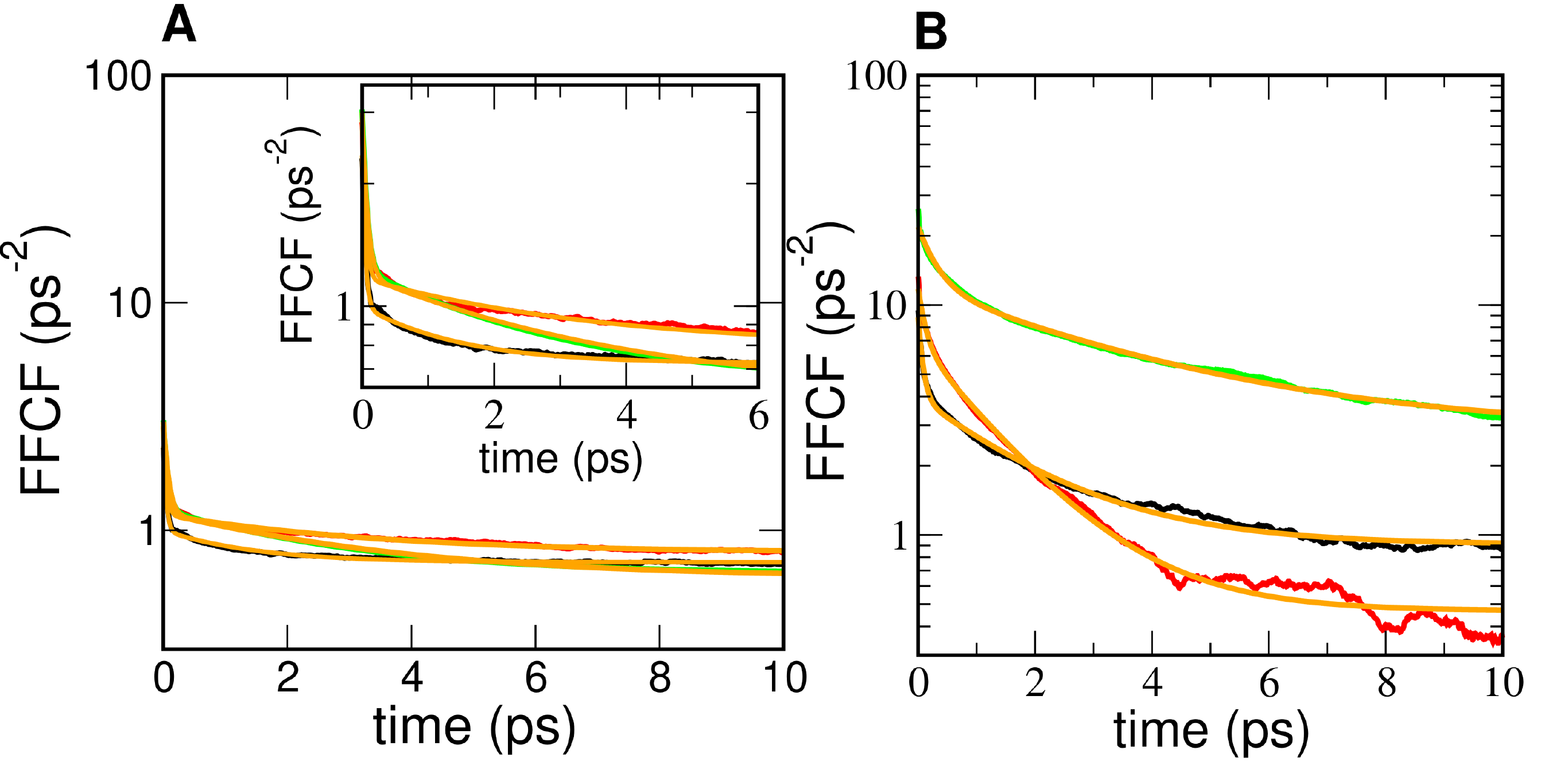}
\caption{FFCF for each -CO moiety of trialanine using the MTP model
  (panel A) and from SCC-DFTB simulations (panel B) for central -CO
  (black), outer -CO (red), and the CO(OH) group (green). The orange
  lines are the fit to Eqn. \ref{eq:fit.ffcf} for each case.}
\label{fig:fig4}
\end{figure}

\noindent
{\it Frequency Fluctuation Correlation Function:} The frequency
fluctuation correlation function provides information about the
environmental dynamics surrounding a local spectroscopic probe and the
coupling to it. The FFCFs were fitted to a bi-exponential decay with
static component (equation \ref{eq:fit.ffcf}) as has also been done
for NMAD.\cite{stock.nma.jcp.2002} The raw data with the corresponding
fits are shown in Figure~\ref{fig:fig4} and the parameters are
summarized in Table~\ref{tab:tab2}. FFCFs for each -CO probe of
trialanine using the MTP model (Figure \ref{fig:fig4}A) and from SCC-DFTB
simulations (Figure \ref{fig:fig4}B) for central (black), outer (red), and
terminal (green) -CO are reported.\\

\begin{table*}[!h]
\centering
\begin{tabular*}{\linewidth}{@{\extracolsep{\fill}} lcccccc}
\hline\hline
 Model & Mode & $a_1$ [ps$^{-2}$] & $\tau_1$ [ps] & $a_2$ [ps$^{-2}$] & $\tau_2$ [ps] & $\Delta_0^2$ [ps$^{-2}$]  \\
\hline
MTP (FNM) & --CO(central) &  1.348 & 0.038 & 0.274 & 1.337 & 0.723 \\                  
&--CO (outer) &1.709  & 0.044 & 0.373 & 3.066 & 0.797 \\      
&--CO(OH) &    1.804 &   0.067 & 0.568 & 3.184 & 0.620  \\    
\hline
 MTP (INM) & --CO (central) & 2.250 & 0.057 & 0.662 & 1.043  & 0.087 \\
                          & --CO(outer) & 2.360 & 0.076 & 0.419 & 1.543 & 0.023 \\
\hline
PC (FNM) &--CO (central) & 1.391 & 0.039 & 0.204 & 2.306 & 0.200 \\
&--CO (outer) & 1.942 & 0.037 & 0.422 & 4.697 & 0.419 \\
&--CO(OH) & 1.465 & 0.098 & 1.350 & 6.028 & 0.099 \\
\hline
SCC-DFTB (FNM) &-CO(central) & 4.872 & 0.076 & 3.052 & 1.845 & 0.908 \\
&-CO (outer) & 4.981 & 0.124 & 6.299 & 1.350 & 0.468 \\
&-CO(OH) & 9.780 & 0.311 & 9.198 & 3.508 & 2.877 \\
\hline
\end{tabular*}
\caption{Parameters for fitting the FFCFs to a bi-exponential decay
  with static component fit (Eqn. \ref{eq:fit.ffcf}) for trialanine
  from simulations using the PC, MTP, and SCC-DFTB models.}
\label{tab:tab2}
\end{table*}

\noindent
The short
time scale $\tau_1$ ranges from 0.04 to 0.07 ps whereas the longer one
ranges from 1.3 ps to 3.2 ps. Using PC simulations the decay time
$\tau_1$ is similar to that from the MTP simulation whereas the long
time scales increase by about a factor of two. The amplitudes ($a_1$
and $a_2$) of the two time scales are comparable for the two
methods. For the simulation with MTPs, the static components for the
central and outer amide are similar in magnitude, on average
$\Delta_0^2 \sim 0.75$ ps$^{-2}$ (which yields $\Delta_0\sim 0.866$ ps$^{-1}$ equivalent to $\Delta_0 = 4.6$
cm$^{-1}$), which is in good agreement with the experimentally
reported value\cite{Woutersen} of $\Delta_0 = 5$ cm$^{-1}$. This
static component appears for (Ala)$_3$ but not for NMA and is
quantitatively captured by using the MTP force field together with the
FNM analysis and consistent with experiment which report that ``In
contrast to NMA, the amide I band of trialanine is still notably
inhomogeneous on the 4 ps time scale.''\cite{Woutersen} For the
simulations with the PC model, the fits to Eq. \ref{eq:fit.ffcf} yield
$\Delta_0^2 = 0.20$ ps$^{-2}$ and $\Delta_0^2 = 0.42$ ps$^{-2}$ which
is smaller by about a factor of two compared with experiment. Also,
the two static components for the central and outer -CO label differ
by a factor of two.\\

\noindent
With SCC-DFTB, the short time decay $\tau_1$ is considerably slower
(0.1 ps to 0.3 ps) and the longer time scales range from 1.4 ps to 3.5
ps. The short time decay is considerably longer than that reported
from experiment whereas the long time decay for the central and outer
-CO are compatible with $\tau_c = 1.6$ ps used for interpreting
experiments on (Ala)$_3$ which was, however, fixed at the value found
for NMA.\cite{Woutersen} The values of $\Delta_0^2$ for the central
and outer -CO differ by a factor of two, similar to the results from
the simulations with PCs but on average, they are consistent with the
experimental value.\cite{Woutersen}\\

\noindent
The magnitude of $C(t=0)$ (i.e. the FFCF at $t=0$) has been reported
to be $\Delta_1^2 = 121$ cm$^{-2}$ equivalent to 4.30
ps$^{-2}$.\cite{stock.nma.jcp.2002} This compares with values of 1.65
ps$^{-2}$, 2.05 ps$^{-2}$, and 2.39 ps$^{-2}$ from simulations with
MTP and 8.5 ps$^{-2}$, 12.78 ps$^{-2}$, and 23.33 ps$^{-2}$, from the
SCC-DFTB simulations for the central, outer and CO(OH) groups. Hence,
the MTP simulations underestimate the experimentally reported
amplitude whereas SCC-DFTB simulations overestimate it by about a
factor or two. This was also found for simulations and experiments on
fluoro-acetonitrile.\cite{facn} The value $C(t=0)$ is a measure of the
interaction strength between the reporter(s) and the
environment. Thus, the present findings suggest that this interaction
is underestimated by the MTP model and overestimated by SCC-DFTB. Such
information can be used to further improve the energy function.\\

\noindent
Considering the results on the FFCFs for NMAD and (Ala)$_3$ together
is it noted that only the simulations with MTP are consistent with
experiment in that a) their decay times are close to one another and
b) the fact that the FFCF for NMAD has no static component but that
for (Ala)$_3$ has $\Delta_0^2 > 0$. It is also of interest to note
that the fast decay time $\tau_1 \lesssim 100$ fs of the FFCF observed
in the present simulations is consistent with an experimentally
observed time constant of $\tau = 110 \pm 20$
fs.\cite{Hamm.pnas.2001}\\

\noindent
The associated lineshapes for the three different modes involving the
-CO stretch for trialanine are calculated via 1D Fourier
transformation of the lineshape function as was done for NMAD, see
Figures~\ref{fig:fig3}C and D. The FWHM for the 1D-IR spectra are
13\,cm$^{-1}$ for the central -CO, 17\,cm$^{-1}$ for the outer one and
18\,cm$^{-1}$ for the terminal -CO(OH) using the MTP model and
25\,cm$^{-1}$, 32\,cm$^{-1}$, and 50\,cm$^{-1}$ when using the
SCC-DFTB model. Experimentally,\cite{stock.nma.jcp.2002} it was found
that the FWHM for NMAD and (Ala)$_3$ differ little and are $\sim 20$
cm$^{-1}$. Both findings are quite well captured by the MTP
simulations whereas with SCC-DFTB the widths are larger and differ
somewhat more between NMAD and (Ala)$_3$.\\

\begin{figure}[H]
\includegraphics[width=15cm]{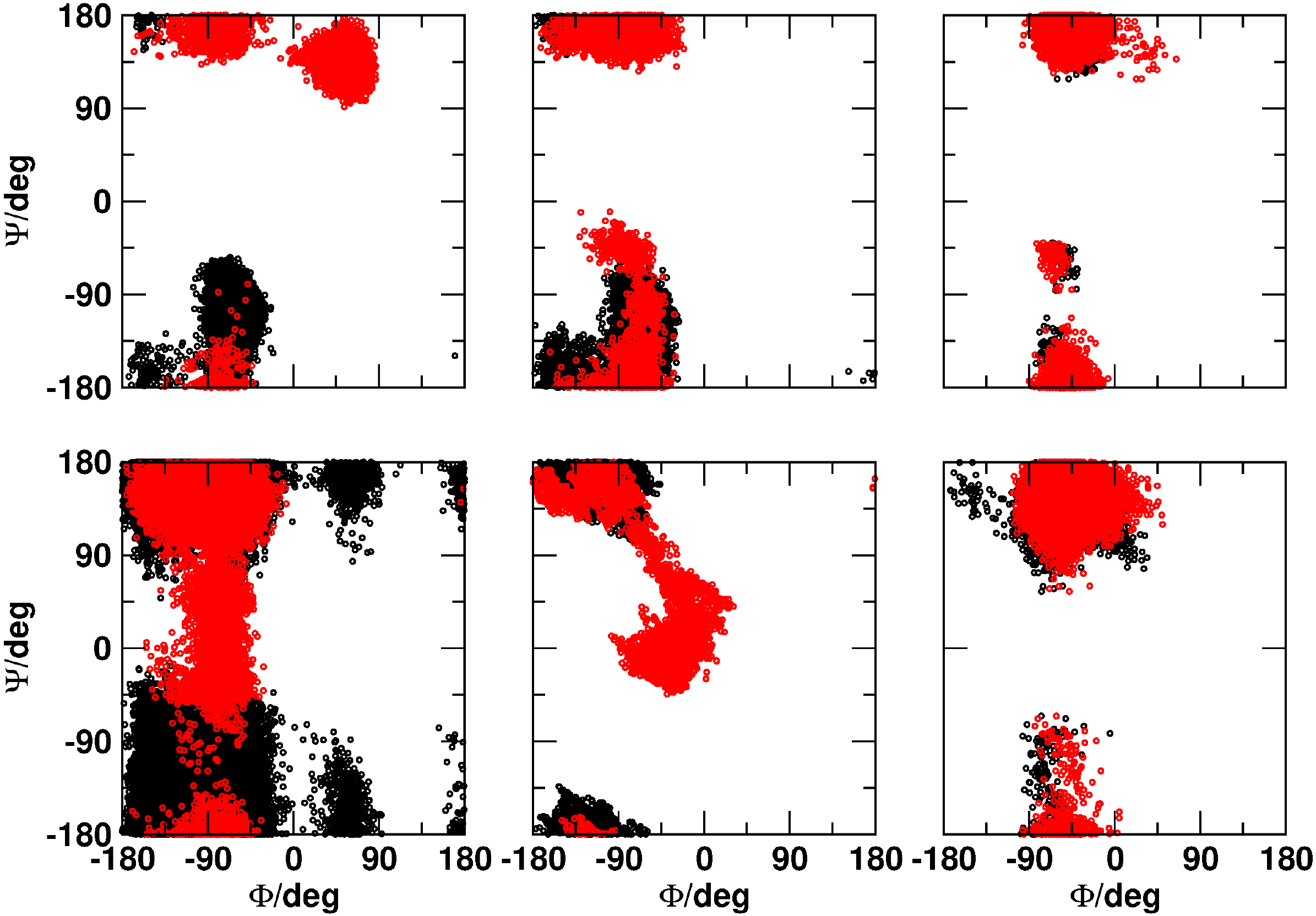}
\caption{Ramachandran plots ($\Phi/\Psi$ angle) of trialanine using
  the PC (left panel), SCC-DFTB (middle panel) and MTP (right panel)
  models. Top and bottom panels are from simulations in the gas-phase
  and in water, respectively. Black are the Ramachandran angles, and
  red are the $\Phi$/$\Psi$ angles of trialanine for the carboxylic
  terminus. The centers for the $[\Phi,\Psi]$ angles for the $\beta$,
  P$_{\rm II}$, $\alpha_R$ and $\alpha_L$ conformations are
  $[-140^{\circ},130^{\circ}]$, $[-75^{\circ},150^{\circ}]$,
  $[-70^{\circ},-50^{\circ}]$, and $[50^{\circ},50^{\circ}]$,
  respectively.}
\label{fig:rcplot}
\end{figure}

\noindent
{\it Structural Dynamics:} To characterize the structural dynamics
afforded by the different energy functions used in the present work,
the distribution of $\Phi$/$\Psi$ angles (Ramachandran plot) were
determined from trajectories with the PC, MTP, and SCC-DFTB models,
see Figure \ref{fig:rcplot}. This is used to determine whether,
depending on the energy function used, the conformational space
sampled differs. Also, assessing differences in the sampling between
simulations in the gas phase and in solution are of interest. Both,
the conventional $[\Phi,\Psi]$ map involving the central and outer -CO
labels, and the dihedral angles for the terminal -CO are reported.\\

\noindent
Figure \ref{fig:rcplot} shows the Ramachandran plot for trialanine
from simulations using the PC (left panel), SCC-DFTB (middle panel)
and MTP (right panel) models. The centers for the $[\Phi,\Psi]$ angles
for the $\beta$, P$_{\rm II}$, $\alpha_R$ and $\alpha_L$ conformations
are $[-140^{\circ},130^{\circ}]$, $[-75^{\circ},150^{\circ}]$,
$[-70^{\circ},-50^{\circ}]$, and $[50^{\circ},50^{\circ}]$,
respectively. From simulations in the gas phase (top) the
distributions for the regular Ramachandran angles from PC and SCC-DFTB
simulations are similar. They both sample $\beta$, P$_{\rm II}$, and
$\alpha_{\rm R}$ structures. For simulations with MTP the densities
are somewhat more shifted towards the P$_{\rm II}$ structures and the
$\alpha_R$ state is sampled as well. For the COOH group (red), the
region for $\Phi >0$ is occupied for simulations with PCs but not with
SCC-DFTB. For simulations with the MTP model the same regions as for
the regular Ramachandran angles are sampled.\\

\noindent
The distribution of conformational state population in Figure
\ref{fig:rcplot} find increased flexibility of (Ala)$_3$ from
simulations with the PC model compared with those using MTP and
SCC-DFTB both in the gas phase and in water. For the simulations in
water (bottom row in Figure \ref{fig:rcplot}), the changes compared
with the gas phase are most pronounced with PCs. In addition to the
$\beta$, P$_{\rm II}$, and $\alpha_{\rm R}$ structures, the poly-Gly
regions are also accessed extensively. Contrary to that, the
differences between the gas and the condensed phase from simulations
with SCC-DFTB and MTP are smaller but nevertheless exhibit increased
flexibility as was found for the simulations with PCs. Using SCC-DFTB
sampling of the $\beta$ and P$_{\rm II}$ structures is extensive
whereas $\alpha_{\rm R}$ is not sampled at all for the regular
Ramachandran angle (but for the -COOH terminus, see red
distribution). Finally, for MTP, the distributions in the region of
the $\beta$ and P$_{\rm II}$ states broaden and there is also some
limited sampling of the $\alpha_{\rm R}$ helix. Both, SCC-DFTB and MTP
only sample ``allowed'' regions in solution whereas PC also accesses
``unusual'' (poly-Gly) and ``forbidden'' regions.\\

\noindent
Ramachandran maps have also been reported from simulations using a
range of parametrized, PC-based force fields, including C27, C36, and
C36m together with the TIP3P and SPC/E water
models.\cite{Tokmakoff2018} The distributions found in the present
work, see Figure \ref{fig:rcplot} lower left panel, are consistent
with these $(\Phi,\Psi)$ maps. Using a Bayesian refinement on the
measured and computed 1d-IR spectra, a consensus 2-dimensional
potential of mean force (PMF) as a function of $(\Phi,\Psi)$ was
determined. Notably, the refined PMF$(\Phi,\Psi)$ vis-a-vis experiment
reported in Ref.\cite{Tokmakoff2018} closely resembles the
distribution found from the MTP simulations, see black symbols in
Figure \ref{fig:rcplot} lower right panel.\\

\noindent
Table \ref{tab:tabconf} summarizes state populations for $\beta$,
P$_{\rm II}$, $\alpha_R$ and $\alpha_L$ conformations of trialanine
from simulations with PC, MTP, and SCC-DFTB models. A comparison with
several previous studies is also
provided.\cite{woutersen:2000,Schweitzer-Stenner.jacs.2001,Mu.jpcb.2002,Graf2007,Oh2010,Xiao2014,Tokmakoff2018}
For assigning a particular conformation to one of the 4 states, first
centers ($\Phi$, $\Psi$) of each of the states were defined as
$[-140^{\circ},130^{\circ}]$, $[-75^{\circ},150^{\circ}]$,
$[-70^{\circ},-50^{\circ}]$, and $[50^{\circ},50^{\circ}]$ for
$\beta$, P$_{\rm II}$, $\alpha_R$ and $\alpha_L$ conformations,
respectively. If a particular conformation is within $\pm 40^\circ$
around any of the centers, the conformation is assigned to that
center. If a conformation is outside these bounds, it is not assigned
which is the case for 30 \% to 40 \% of the structures. Then the
percentage for the population of a particular substate was determined
as the fraction of all assigned conformations. The present simulations
using MTP find dominant population of the P$_{\rm II}$ state (98 \%)
with a small fraction of $\beta$ and $\alpha_R$. Using a PC model,
P$_{\rm II}$ is still most populated, followed by $\beta$ and
$\alpha_{\rm R}$. Simulations with SCC-DFTB yield a higher population
of $\beta$, a smaller fraction for P$_{\rm II}$ and no helical
conformations.\\

\noindent
Most previous studies find that the P$_{\rm II}$ state is most
populated, typically followed by $\beta$ structures. The relative
populations range from 66 \% to 92 \% for P$_{\rm II}$ and 0 to 23 \%
for $\beta$. Fewer studies report population of $\alpha_{\rm R}$. One
of the most sophisticated investigations (Bayesian ensemble refinement
against FTIR and 2DIR experimental data)\cite{Tokmakoff2018} report a
$(85 \pm 6)$\% population for P$_{\rm II}$, $(14 \pm 5)$\% for $\beta$
and an insignificant population $(1 \pm 2)$\% for $\alpha_{\rm
  R}$. Within error bars, the results from the MTP simulations are
consistent with these findings. It is interesting to note that the
``original'' state populations in the work by Tokmakoff et al. were
all derived from MD simulations using PC-based force fields and the
populations are largely independent on the particular choice of the
all-atom FF, see Table S5 in Ref.\cite{Tokmakoff2018} Specifically,
the populations from the C36 parametrization with the TIP3P water
model (as used here, see Table S4 in Ref.\cite{Tokmakoff2018}) compare
favourably with the present findings for $\beta$ structures (18\%
vs. 20 \%), P$_{\rm II}$ (68 \% vs. 79 \%), and $\alpha_{\rm R}$. (6
\% vs. 3\%). Differences may arise due to slightly different
definitions of the basins to integrate the populations and whether or
not all of the conformations are used for analysis. After Bayesian
refinement the populations are comparable to those from MTP
simulations. In other words, machine learning of the populations based
on the comparison of measured and computed IR spectra has the same
effect as replacement of PCs by MTPs in the present simulations,
lending additional support to the physical relevance afforded by the
anisotropic effects in the electrostatic interactions.\\

\begin{table*}[!h]
\centering
\begin{tabular}{ccccc}
\hline\hline
\multicolumn{5}{c}{ Conformational State Population}  \\
 \hline
                                 & $\beta$ & P$_{\rm II}$ & $\alpha_R$ & $\alpha_L$   \\
 \hline 
MTP (this work)       &     1\%            &        98\%           &    1\%                &     0   \\    
PC (this work) &           18\%    &             79\%           &      3\%    &            $<$1\%  \\
SCC-DFTB (this work)     &                      62\%      &           38\%      &            0      &                0 \\
\hline
Tokmakoff et al. (original) \cite{Tokmakoff2018}$^{,a}$              & $(22\pm7)$\% & $(63\pm11)$\%) & $(12\pm8)$\% & $(3\pm2)$\% \\
Tokmakoff et al. (Refined)$^{b}$                                                      & $(14\pm5)$\% & $(85\pm6)$\% & $(1\pm2)$\% & $<0.1$\% \\
Woutersen et al.\cite{woutersen:2000}$^{,c}$                                      & 0\% & 80\% & 20\% & 0\% \\
Mu et al.\cite{Mu.jpcb.2002}                                      & 42\% & 41\% & 16\% & 0.8\% \\
Schweitzer-Stenner  \cite{Schweitzer-Stenner.jacs.2001}$^{,d}$ & 16\% & 84\%& 0\% & 0\% \\
Graf et al.\cite{Graf2007}$^{,e}$                                                     & 8\% & 92\%& 0\% & 0\%  \\
Oh et al. \cite{Oh2010}$^{,f}$                                                         & 12\% & 88\% & 0\% & 0\% \\
Xiao et al. \cite{Xiao2014}$^{,g}$                                                   & $2.0\pm1.8)$\%   & $(85.8\pm4.9)$\% &   $(5.5\pm4.1)$\%  &  $(3.5\pm2.7)$\% \\
Beauchamp et al.  \cite{Beauchamp2014}$^{,h}$                         & $(23\pm6)$\% & $(67\pm9)$\% & $(10\pm8)$\% & -- \\
\hline
\end{tabular}
\caption{State population for $\beta$, P$_{\rm II}$, $\alpha_R$ and
  $\alpha_L$ conformations of trialanine from simulations with PC,
  MTP, and SCC-DFTB compared with previously reported values. $^a$MD
  simulation, $^b$Bayesian ensemble refinement against FTIR and 2D IR,
  $^c$Fitting 2D IR spectra, $^d$Fitting VCD, Raman, FTIR and
  J-coupling, $^e$Fitting NMR, $^f$Fitting NMR with Gromos 43A1,
  $^g$Fitting NMR with integrated Bayesian approach, $^h$Bayesian
  energy landscape tilting. The standard deviation from the average is
  given in parentheses.}
\label{tab:tabconf}
\end{table*}

\subsection{FFCF from Independent Normal Modes} 
The amide-I vibrational dynamics encoded in the FFCF contains
information about the solvent dynamics as well as the peptide
conformational dynamics. To better understand the influence of
inter-mode couplings on the conformational dynamics, the instantaneous
normal modes for the central and the outer -CO label were also
determined from normal mode analyses treating the two amide modes
independently. This is then compared with the FFCFs obtained from the
FNM analysis which contains the couplings between the labels.\\

\begin{figure}[H]
\includegraphics[width=10cm]{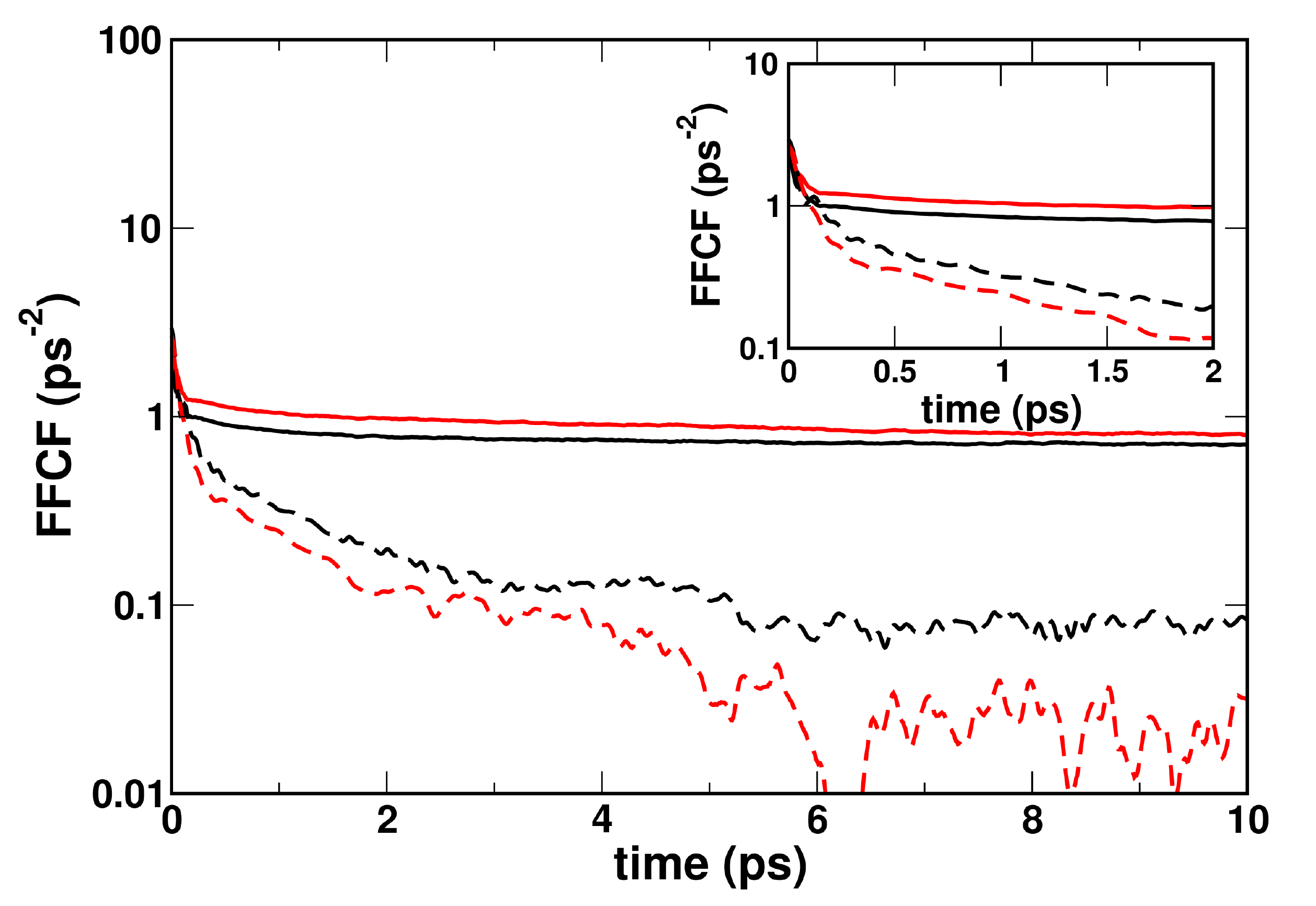}
\caption{Comparison of FFCFs from full NM analysis (solid lines) and
  independent NM analysis (dashed lines) for the outer (red) and
  central (black) amide modes of trialanine.}
\label{fig:compare_full_part}
\end{figure}

\noindent
The FFCFs for the central and the outer -CO from INM (dashed lines)
and from FNM (solid lines) are reported in Figure
\ref{fig:compare_full_part} and the fitting parameters to
Eq. \ref{eq:fit.ffcf} using two time scales are given in Table
\ref{tab:tab2}. Without coupling (dashed lines) the FFCFs decay close
to zero on the 10 ps time scale and the magnitude of $\Delta_0^2$
decreases by almost one order of magnitude compared with the results
from FNM. Also, the decay times are shorter if the coupling between
the two labels is neglected. As the results from FNM analysis agree
with experiment and those omitting the coupling do not, it is
concluded that the FNM analysis together with a MTP representation of
the electrostatics provides a means to correctly describe the dynamics
of hydrated (Ala)$_3$.\\

\noindent
Including couplings between the labels (``sites'') is also important
when working with map-based approaches for 1d- and 2d-IR
spectroscopy.\cite{knoester-jpc-model-2006,skinner-map-JPCB2011,Tokmakoff-map-jcp-2013}
Using frequency maps, the site energies, the nearest neighbor coupling
and the transition dipole couplings are usually included in the
excitonic Hamiltonian.\cite{skinner-map-JPCB2011} Such couplings need
to be (re-)introduced in an excitonic Hamiltonian but they are already
partly present in the FNM approach used here, as the above analysis
demonstrates. The molecular dynamics simulations which generate the
conformational ensemble to be analyzed include couplings through the
nuclear dynamics and the FNM analysis preserves these couplings
whereas the INM analysis almost entirely removes them.\\

\noindent
Comparing the maxima of the peak positions from the frequency
distributions based on ``full NM'' and ``independent NM'' reveals that
the two analyses differ in capturing this coupling. From INM the
frequency distributions peak at 1661.5 cm$-1$ and 1662 cm$^{-1}$,
i.e. a splitting of close to zero, whereas from FNM the maxima are at
1670 cm$^{-1}$ and 1683 cm$^{-1}$, i.e. a splitting of 13
cm$^{-1}$. Within a simple two-state Hamiltonian this amounts to a
coupling of $\sim 6.5$ cm$^{-1}$, consistent with
experiments.\cite{Hamm.pnas.2001}\\

\noindent
The finite amplitude of $\Delta_0^2$ is also indicative of the fact
that within the explored time scale the system has not exhaustively sampled
all available states. In other words, population
relaxation is not complete on the 10 ps time scale. This is consistent
with an analysis of MD trajectories that determined the FFCF from only
sampling the P$_{\rm II}$ conformation (which decays to zero on the
$\sim 4$ ps time scale) compared with the full MD trajectory sampling
different substates for which a static contribution remains even after
10 ps.\cite{stock.nma.jcp.2002} This interpretation is also consistent
with the fact that NMAD only has one conformational substate and
therefore the FFCF decays to zero on the 10 ps time scale.

\section{Summary and Conclusion}
In summary, the present work provides a comprehensive assessment and
comparison of the dynamics and infrared spectroscopy of NMAD and
(Ala)$_3$ in D$_2$O. Consistent with experiments on (Ala)$_3$ it is
found that with ``full normal modes'' from simulations using MTPs to
compute the frequency trajectory, the 1d-infrared spectrum for the
outer and central -CO labels are split by 13 cm$^{-1}$, compared with
25 cm$^{-1}$ from experiment. With independent normal modes this
splitting is close to zero. Including the site-site couplings in the
NM analysis therefore yields a more quantitative description of the
spectroscopy and dynamics. This splitting is larger (47 cm$^{-1}$)
in simulations with SCC-DFTB for the solute. The FFCF from FNM has an
initial amplitude $C(t=0)$ of [1.65, 2.05] ps$^{-2}$ for the central
and outer -CO label, compared with 4.30 ps$^{-2}$ from experiment and
[8.5, 12.78] ps$^{-2}$ from simulations with SCC-DFTB. This points
towards somewhat weaker interactions of the -CO labels with the
environment in the MTP simulations and a considerably stronger
interaction in SCC-DFTB. The long-time static component from MTP
simulations with FNM of $\Delta_0 = 4.6$ cm$^{-1}$ compares well with
that observed experimentally ($\Delta_0 = 5.0$ cm$^{-1}$) whereas that
from simulations with PCs is smaller by a factor of two. The MTP
simulations find comparable values for $\Delta_0$ for the central and
outer -CO whereas with SCC-DFTB they differ by about a factor of two
with one of the values $\sim 20$ \% larger than that observed
experimentally and the other one lower by a similar amount, see Table
\ref{tab:tab2}.\\

\noindent
Overall, simulations for (Ala)$_3$ with MTP and FNM analysis find good
to quantitative agreement with experiment for the splitting, amplitude
of $C(t=0)$, and value for $\Delta_0$. This contrasts with simulations
using PC and/or INM or SCC-DFTB simulations. The conformational space
sampled by (Ala)$_3$ in solution is dominated by a P$_{\rm II}$
structure (98 \%), followed by $\beta$ and $\alpha_R$, each populated
in 1 \% of the cases. This agrees qualitatively with a Bayesian
refined analysis\cite{Tokmakoff2018} of recent infrared experiments
which find occupations of [P$_{\rm II}$, $\beta$, $\alpha_R$] [$85 \pm
  6$, $14 \pm 5$, $1 \pm 2$]\% but differ somewhat from earlier
results\cite{woutersen:2000} which report [80, 0, 20]\%.\\

\noindent
The present work demonstrates that the structural dynamics of a small,
hydrated peptide can be correctly described from MD simulations based
on an MTP force field in explicit solvent together with a full normal
mode analysis. Such studies provide the necessary basis to link
structural dynamics, spectroscopy and aggregation in larger proteins
from experiment and simulations.\\

\section{Acknowledgments}
The authors gratefully acknowledge financial support from the Swiss
National Science Foundation through grant 200021-117810 and to the
NCCR-MUST. The authors thank Prof. Peter Hamm for valuable
discussions. \\


\end{document}